\documentclass[conference]{IEEEtran}
\IEEEoverridecommandlockouts

\usepackage{cite}
\usepackage{amsmath,amssymb,amsfonts}
\usepackage{algorithmic}
\usepackage{graphicx}
\usepackage{textcomp}
\usepackage{xcolor}
 \usepackage{booktabs} 
 \usepackage{tabularx}    
\usepackage{ragged2e}    
  \usepackage{url}
\def\BibTeX{{\rm B\kern-.05em{\sc i\kern-.025em b}\kern-.08em
    T\kern-.1667em\lower.7ex\hbox{E}\kern-.125emX}}

 \usepackage{geometry}
\begin{document}

\title{Policy-Driven AI in Dataspaces: Taxonomy, Explainability, and Pathways for Compliant Innovation}

\author{\IEEEauthorblockN{Joydeep Chandra}
\IEEEauthorblockA{\textit{Department of CST} \\
\textit{Tsinghua University}\\
Beijing, China \\
joydeepc2002@gmail.com}
\and
\IEEEauthorblockN{Satyam Kumar Navneet}
\IEEEauthorblockA{\textit{Department of CSE} \\
\textit{Chandigarh University}\\
Mohali, India\\
navneetsatyamkumar@gmail.com}}

\maketitle
\begin{abstract}
As AI-driven dataspaces become integral to data sharing and collaborative analytics, ensuring privacy, performance, and policy compliance presents significant challenges. This paper provides a comprehensive review of privacy-preserving and policy-aware AI techniques, including Federated Learning, Differential Privacy, Trusted Execution Environments, Homomorphic Encryption, and Secure Multi-Party Computation, alongside strategies for aligning AI with regulatory frameworks such as GDPR and the EU AI Act. We propose a novel taxonomy to classify these techniques based on privacy levels, performance impacts, and compliance complexity, offering a clear framework for practitioners and researchers to navigate trade-offs. Key performance metrics--latency, throughput, cost overhead, model utility, fairness, and explainability--are analyzed to highlight the multi-dimensional optimization required in dataspaces. The paper identifies critical research gaps, including the lack of standardized privacy-performance KPIs, challenges in explainable AI for federated ecosystems, and semantic policy enforcement amidst regulatory fragmentation. Future directions are outlined, proposing a conceptual framework for policy-driven alignment, automated compliance validation, standardized benchmarking, and integration with European initiatives like GAIA-X, IDS, and Eclipse EDC. By synthesizing technical, ethical, and regulatory perspectives, this work lays the groundwork for developing trustworthy, efficient, and compliant AI systems in dataspaces, fostering innovation in secure and responsible data-driven ecosystems.
\end{abstract}

\begin{IEEEkeywords}
Dataspaces, Privacy-Preserving AI, Policy-Aware AI, Explainable AI (XAI), Privacy-Performance Trade-offs
\end{IEEEkeywords}

\section{Introduction}\label{sec1}
The usage of Artificial Intelligence (AI) systems in many areas has changed how data is utilized in  diverese sectors, especially in emerging dataspace\cite{ds} architectures. As AI becomes more common in these areas, a balance between data privacy and the need for performance in data processing emerges. This balance calls for a close look at existing plans that tried to meet these competing needs. Recent studies showed a change in how firms see privacy work, noting that 96\% of people saw more benefits than costs\cite{cisco2025}. Also, 88\% of firms in 2025 said that building trust in their brand was a big reason for their privacy efforts\cite{electronica}. The above data shows how strong privacy measures helps build trust with customers, which then helps businesses grow and retain their customers, changing how firms use AI in dataspaces\cite{ds}. Also, new regulatory governance in organizations around the world made privacy more critical, turning it into a responsible AI development and business growth\cite{Solove2025}. 

As more people got used to Generative AI (GenAI), as shown by 63\% of respondents in 2025, was worried about sharing sensitive data\cite{cisco2025}. This tension underlines a key point: AI needs big data sets for training\cite{zhao2025surveylargelanguagemodels}, which raises risks about privacy. Because of this, firms planned to move resources from privacy budgets to AI initiatives, with IT spending expected to nearly double\cite{cisco2025}. This shows that the need for data in AI requires stronger privacy enhancements, which costs more money, often taking from current privacy funds\cite{Solove2025}. \textbf{We make the following contributions through this survey in the field of AI-driven dataspaces:}
\begin{itemize}
    \item Comprehensive review of privacy-preserving AI techniques, including Federated Learning\cite{fed}, Differential Privacy\cite{dps}, Trusted Execution Environments\cite{7345265}, Homomorphic Encryption\cite{homo}, and Secure Multi-Party Computation, analyzing their privacy guarantees, performance impacts, and compliance complexities was provided.
    \item A novel taxonomy that categorizes these techniques based on privacy levels, performance degradation, and regulatory alignment, offering a structured framework for researchers and practitioners to navigate trade-offs was proposed.
    \item Critical research challenges, such as the lack of standardized privacy-performance metrics, explainability in federated ecosystems, and semantic policy enforcement were exposed.
    \item Future directions were outlined, proposing a conceptual framework for policy-driven AI alignment, automated compliance validation, standardized KPIs, and integration with European initiatives like GAIA-X\cite{gaia}, IDS\cite{ids}, and Eclipse EDC\cite{arnold2025servicearchitecturedataspaces}, to foster trustworthy, scalable, and compliant AI systems in dataspaces.
\end{itemize}

\section{Background}
\subsection{Dataspaces Architecture and AI Integration}
Dataspaces\cite{ds} were a decentralized, open set of rules and frameworks made to let trusted data share among many users and organizations, even among competitors, while ensuring ownership and autonomy\cite{enable}. Unlike centralized platforms, dataspaces\cite{ds} worked as easy parts for any data group, helping trust in data transanctions backed by ways to check and keep trust. This setup let users keep full say over how, when, and with whom data was shared, and under what terms, including adhereing to legal rules.\cite{trustees} The way to technical access in to dataspaces\cite{ds} was built on common rules, making sure different organizations can use and work together. Putting AI into these data spaces brought good and bad points. AI, needing lots of data to learn and get better, used the shared data from dataspace well\cite{zhao2025surveylargelanguagemodels}. But, bringing AI in raised worries about misusing data, how much and why data was gathered, and the hidden ways of many AI systems. Groups in Europe, like GAIA-X\cite{gaia} and the International Data Spaces, made plans to handle these data areas, focusing on data ruling, privacy, secret-keeping, safety, and working well with others. GAIA-X made a clear line between Federations (free groups of people) and dataspace (spaces with certain data swap services like plugs and rules for using data). The Eclipse Dataspace Components offered a full setup, giving main parts like Plugs, Shared Catalogs, and ID Centers to help safe and checked data sharing, backed by Gaia-X\cite{gaia} and IDSA rules\cite{enable}.  

Dataspaces\cite{ds} were designed to enhance data sharing and enable new business models by balancing trust and data autonomy\cite{trustees} which made them ideal environments for AI, which required vast amounts of data for training.\cite{zhao2025surveylargelanguagemodels} However, by enabling more extensive data flows, dataspaces\cite{ds} increased the existing challenges of data privacy and security issues, as sensitive information could be more widely distributed. The mechanism which was initially designed to enahance data's potential for AI also increased the attack surface and  the complexity of maintaining privacy. This indicated that privacy-by-design principles were not merely beneficial but essential for dataspace\cite{ds} architectures from their inception, rather than being an afterthought. The adoption of dataspaces\cite{ds} rested on standardized rules to make sure tech could work together and allow applications in different kinds of setup. Projects like GAIA-X\cite{gaia} and IDS\cite{ids} really aimed to set even rules and ways to check trust\cite{trustees}. The Eclipse Dataspace Components (EDC)\cite{arnold2025servicearchitecturedataspaces} made this clearer by giving a rule-based way for builders. This showed that without open, agreed rules and checkable ways to follow them, the hope for trusted, working AI-driven dataspaces could not come true. The big meaning here was that rule groups and big work teams had to put first and speed up making and using these standards to help good AI use. 

\subsection{Fundamentals of Privacy-Preserving Computation}
Privacy-Preserving Computation (PPC)\cite{ppc} uses many ways to do computation on private data without showing the raw informnation. These ways are key when AI applications operating in dataspaces\cite{ds} where keeping data safe and following rules are very important. Key PPC\cite{ppc} techniques are Federated Learning (FL)\cite{fed}, Differential Privacy (DP)\cite{dps}, Trusted Execution Environments (TEEs)\cite{7345265}, Homomorphic Encryption (HE)\cite{homo}, and Secure Multi-Party Computation (SMC)\cite{smc}. FL lets AI models learn together from lots of data spots, like phones or places, without putting the raw data all in one spot\cite{mahlool2022comprehensivesurveyfederatedlearning} . This stops the need to give out raw user data to a big server, cuts down costs, and keeps data more private.\cite{computers13110277} . Here, local devices work on models with their own data and send back just the model changes to a big server to mix together. This mechanism is essential to apply in areas such as banks or hospitals, where keeping data private and specific rules to make it protected from exploitation is very important.\cite{mahlool2022comprehensivesurveyfederatedlearning}.

DP further enhanced privacy in FL by introducing carefully calibrated noise into model updates, making it mathematically difficult to infer individual data points.\cite{computers13110277} TEEs offered hardware-based isolation, creating secure confidential computing spaces where code and data remained encrypted and inaccessible to external software or operating systems, enabling faster inference compared to purely cryptographic methods.\cite{Goretti_2024} HE allowed computations directly on encrypted data without prior decryption, ensuring confidentiality throughout processing.\cite{10.1145/3715877} SMC\cite{smc} enabled multiple parties to jointly compute functions on their private inputs without revealing the inputs themselves.\cite{Gamiz2025} Each of these techniques presented distinct trade-offs between privacy guarantees, computational efficiency, scalability, and practical applicability, which were critical considerations for their deployment in AI-driven dataspaces\cite{ds}.

The review of PPC\cite{ppc} techniques revealed a spectrum of approaches, from FL and DP (software-based, often introducing utility trade-offs) to TEEs (hardware-based, offering efficiency but with specific trust assumptions) and HE\cite{9734024}/SMC\cite{smc} (cryptographic, providing strong guarantees but with high computational overhead).\cite{computers13110277} This indicated that no single technology was universally optimal. Instead, a hybrid approach, combining different PETs, was often necessary to achieve a balanced privacy-performance profile.\cite{Gamiz2025} For example, FL could be combined with DP for stronger privacy \cite{mahlool2022comprehensivesurveyfederatedlearning}, or HE could be integrated with SMC\cite{smc} for enhanced security and efficiency in complex BI workloads.\cite{Gamiz2025} This suggested that future solutions would increasingly involve multi-layered PET architectures tailored to specific use cases and threat models within dataspaces\cite{ds}. While FL was initially seen as a strong privacy-preserving method by keeping raw data local, recent studies demonstrated that private data could still be inferred from uploaded model parameters.\cite{10.1145/3700838.3703679} This highlighted a subtle but critical point: privacy was not merely about preventing direct data access but also about mitigating inference attacks and data leakage from indirect information. The evolution of privacy-preserving techniques was accompanied by the sophistication of potential attacks, necessitating continuous refinement of privacy guarantees (e.g., through DP noise injection \cite{computers13110277}). This implied a dynamic and adversarial landscape where the "balance" between privacy and performance was a moving target, requiring ongoing research and adaptive strategies.

\section{Policy-Aware AI Alignment}
\subsection{Regulatory Landscape and Compliance Frameworks}
Worldwide acknowledgment of the need for robust AI and data privacy regulations transformed the legal landscape between 2020 and 2025. Key frameworks, particularly in the European Union, established strict rules for AI systems managing sensitive data in dataspaces \cite{ds}. The General Data Protection Regulation (GDPR) continued to be a central pillar, overseeing the collection, processing, and storage of personal data while requiring explicit user consent \cite{SAURA2022101679}. GDPR stressed data minimization and purpose limitation, ensuring AI systems only gather and use the data needed for their specific purposes \cite{10.1145/3531532}. Homomorphic Encryption (HE) \cite{homo} was recognized as a method to support GDPR compliance by keeping data encrypted during processing, potentially reducing the need to report data breaches if the data stayed unreadable without the key \cite{10.1145/3715877}.

Kicking off in August 2024 and fully rolling out by August 2026, the EU AI Act brought a new system to rank AI based on risk levels\cite{Enqvist2024}. It sorted AI into four groups: unacceptable, high, limited, and minimal or no risk. Things like harmful AI tricks or social scoring were completely banned, while high-risk AI, used in places like critical infrastructure, schools, hiring, law enforcement, or courts, had to follow strict rules. \cite{veale} These rules included checking and reducing risks, using top-notch data to avoid unfair results, keeping activity logs for tracking, providing detailed records, ensuring human oversight, and making systems tough, secure, and accurate. For limited-risk AI, transparency was the main focus, such as informing individuals they're interacting with AI or indicating AI-grown content \cite{megale}. The EU AI Act also imposed regulations on general-purpose AI models, mandating openness and adherence to copyright legislation, along with further steps to manage risks for those that would constitute major problems \cite{10.1145/3531532}.

The Data Governance Act (DGA) and the Data Act generally reinforced such regulations through data sharing and reuse across sectors and ensuring data control and sovereignty \cite{doi:10.1177/17835917221141324}. These legislations were intended to establish a single, cohesive data market, necessitating trustworthy data exchange systems within dataspaces \cite{ds}. Furthermore, certain usage control policies and models were created. The ODRL Data Spaces (ODS) profile supplemented the ODRL core vocabulary for machine-readable directions on data assets of large data systems \cite{plazaortiz2025authenticationauthorizationdataspaces}. ODS established activities such as Query, Publish, Train, Evaluate, Anonymize, and Transform, and roles such as Provider, Consumer, Controller, and Broker for well-defined permissions and prohibitions to be used for AI compliance \cite{SAURA2022101679}. The International Data Spaces (IDS) \cite{ids} Reference Architecture Model (RAM) also created a technology-neutral IDS Usage Control Language, based on ODRL \cite{plazaortiz2025authenticationauthorizationdataspaces}, to set rules for permissions, prohibitions, and obligations for data use \cite{gaia}. IDS RAM\cite{ids} included 21 policy classes, with the Dataspace Connector applying eight, covering cases like unrestricted use, connector-limited use, time-limited use, and logging requirements \cite{dam}. Enforcing these policies in IDS \cite{ids} involved converting them into technology-specific rules during operation \cite{arnold2025servicearchitecturedataspaces}.

Earlier privacy laws like GDPR focused mainly on protecting data. However, the EU AI Act took a proactive, risk-based approach, requiring specific design and development rules for high-risk AI systems before they reach the market \cite{Enqvist2024}. This marked a major shift from simply responding to data breaches or privacy issues to building compliance and ethical considerations into the entire AI system lifecycle. The complexity and potential societal impact of AI called for a "privacy-by-design" and "ethics-by-design" approach, going beyond basic rule-following to comprehensive governance \cite{zhao2025surveylargelanguagemodels}. This meant organizations needed to create cross-functional AI governance teams and ongoing monitoring processes \cite{10.1145/3531532}. While legal frameworks like GDPR and the AI Act set broad principles, technical frameworks like ODRL Data Spaces and IDS RAM \cite{plazaortiz2025authenticationauthorizationdataspaces, ds, ids} provided machine-readable rules for detailed control over data use \cite{plazaortiz2025authenticationauthorizationdataspaces}. The issue was to ensure that general legal rules were translated into understandable, traceable policies that operate in distributed, complex dataspaces. A mismatch between legal and technical levels would lead to compliance problems or useless policies. The key point is that legal rules and technical systems should cooperate smoothly, and formal verification techniques might be able to guarantee correct alignment without errors \cite{wasil2024verificationmethodsinternationalai}. The Key AI regulatory frameworks proposed or adopted by various regions in the world can be seen in Table \ref{tab:ai_regulatory_frameworks}.

\begin{table*}[htbp]
\centering
\caption{Overview of Key AI Regulatory Frameworks and Usage Control Policies}
\label{tab:ai_regulatory_frameworks}
\begin{tabular}{|p{3cm}|p{3cm}|p{5cm}|p{4cm}|}
\hline
\textbf{Framework/Policy} & \textbf{Scope} & \textbf{Key Provisions (relevant to AI in Dataspaces)} & \textbf{Compliance Implications} \\
\hline
GDPR (General Data Protection Regulation) & EU data protection law & Data minimization, purpose limitation, explicit consent for personal data processing, security of processing, data subject rights \cite{SAURA2022101679} & Requires privacy-by-design, lawful basis for processing, robust security measures, potential breach notification exemptions with strong encryption (HE) \cite{10.1145/3715877} \\
\hline
EU AI Act & EU regulation on AI systems & Risk-based classification (unacceptable, high, limited, minimal/no risk); prohibitions (e.g., social scoring); strict obligations for high-risk AI (data quality, logging, human oversight, robustness, cybersecurity, accuracy); transparency for limited-risk AI; rules for general-purpose AI models \cite{Enqvist2024} & Mandates proactive risk assessment and mitigation; high-quality datasets to prevent bias; detailed documentation and traceability; continuous monitoring; requires explainability for high-risk systems \cite{Enqvist2024} \\
\hline
Data Governance Act (DGA) / Data Act & EU framework for data sharing and reuse & Facilitates trusted data sharing across sectors; ensures data sovereignty and control; promotes data altruism and data intermediaries \cite{doi:10.1177/17835917221141324} & Requires mechanisms for secure and controlled data exchange within dataspaces; promotes interoperability and fair data access \cite{doi:10.1177/17835917221141324} \\
\hline
ODRL Data Spaces (ODS) Profile & Technical policy language for data spaces & Extends ODRL for machine-readable usage control; defines actions (Query, Train, Anonymize, Transform) and party functions (Provider, Consumer, Controller, Broker); specifies permissions/restrictions for data assets \cite{plazaortiz2025authenticationauthorizationdataspaces} & Enables granular, automated policy enforcement; critical for specifying data ownership and usage rights across different phases of AI processing (e.g., ML model training) \cite{plazaortiz2025authenticationauthorizationdataspaces} \\
\hline
IDS Reference Architecture Model (RAM) Usage Control Policies & Technical framework for secure data spaces & Defines technology-independent IDS Usage Control Language (based on ODRL); specifies permissions, prohibitions, obligations (e.g., Allow Usage, Restricted Number of Usages, Local Logging); policy added to resource metadata \cite{ids} & Facilitates organizational and technical usage control enforcement; requires transformation to technology-dependent policies for runtime enforcement; ensures trust and data sovereignty \cite{ids} \\
\hline
\end{tabular}
\end{table*}

\subsection{Techniques for Policy Enforcement}
Enforcing policies in AI-driven dataspaces \cite{ds} meant embedding legal and ethical rules directly into AI model operations and data workflows. Several methods were explored to achieve this, going beyond traditional access control to more flexible, context-aware compliance. Constraint-based optimization was used to ensure AI models stayed within legal and ethical limits \cite{RODGERS2023100925}. This approach put limits on the training process to boost performance in certain cases, considering things like computing power or speed (e.g., FLOPS, latency) \cite{barbara2023neurosymbolicaicompliancechecking}. It could be designed as a problem juggling multiple goals or by including penalties in the model’s loss function \cite{schmitt2024consistencymodelsscalablefast}. For example, in conversational AI systems, user-set constraints helped reduce policy violations in critical business areas while allowing more flexibility in others \cite{zhou2023limaalignment}. A new meta-gradient learning approach was used in large-scale commercial assistants to adjust penalty terms for constraint violations, balancing policy goals with compliance \cite{tunstall2023zephyrdirectdistillationlm}.

Rule-based systems combined with AI were used to enforce governance in predictions and overall AI operations. In the past, rule-based systems were the first step in automating finance tasks, using simple, fixed instructions \cite{PappilKothandapani2025}. As AI and machine learning advanced to predictive analytics and more complex applications, rule-based systems remained useful for compliance monitoring \cite{veale}. AI for financial compliance used machine learning, natural language processing, automation, and generative AI to simplify regulatory tasks, identify risks, and ensure compliance with automated rule checks, smart alerts, and auto-generated audit reports \cite{wu2023bloomberggptlargelanguagemodel}. The goal was to reduce mistakes and make compliance processes more efficient \cite{SAURA2022101679}. Policy injection into AI pipelines involved adding legal rules as strict requirements or penalties during or before model training \cite{10.1145/3531532}. This ensured data protection laws, like requirements for clear consent and data minimization, were built into the system from the start \cite{Solove2025}. For instance, if an AI analyzed customer service interactions, policy injection could ensure personal information was anonymized and customers were informed about how their data was used \cite{electronica}. Strong data security measures, regular audits, and updates to security protocols were crucial for keeping compliance and protecting user privacy throughout the AI pipeline \cite{9194237}.

Previously, verifying AI system compliance typically occurred after they were operational. However, approaches like constraint-based optimization and policy injection\cite{10.1145/3531532} reflect a move toward embedding compliance directly into the development and operation of AI systems. This moves from reacting to problems to designing systems with compliance in mind from the start. The takeaway is that AI development in dataspaces \cite{ds} will likely lean toward DevSecOps-like approaches, where legal and ethical rules are treated like technical must-haves, woven in and tested throughout the AI’s entire lifecycle. Constraint-based optimization often used penalties for breaking rules. \cite{RODGERS2023100925} A new meta-gradient learning method was created to tweak these penalties on the fly, balancing rule-following with practical goals. \cite{zhou2023limaalignment} This points to a key issue: fixed penalties can be too strict or too loose, missing the complex trade-offs in real-world situations. The insight is that dynamic settings and different risk levels call for flexible enforcement that can learn and adjust, rather than rigid rules. This suggests AI systems could, in a way, govern themselves by staying aware of policies, using smart feedback loops and possibly borrowing from adaptive privacy strategies.

\section{Privacy-Preserving AI Methods}
Building AI systems for dataspaces \cite{ds} required strong privacy-preserving technologies (PETs) to protect sensitive data while allowing analysis. Several important methods were thoroughly studied \cite{multi-big}.

\subsection{Federated Learning (FL)}
Federated Learning (FL)\cite{fed}, in which data remains dispersed across various devices or groups, has emerged as a crucial technique for collaboration in machine learning. This reduced communication costs and enhanced data privacy by eliminating the need to send raw data to a central server \cite{mahlool2022comprehensivesurveyfederatedlearning}. \cite{computers13110277} Instead, local devices worked on their own data to train models and only sent updates, like gradients or weights, to a central server to combine them. \cite{mahlool2022comprehensivesurveyfederatedlearning} This method was especially helpful in cases where data was sensitive or rules made it hard to centralize data, such as in healthcare (for example, training models to diagnose diseases across hospitals without sharing patient information) and finance (like detecting fraud or assessing credit across banks). 
Even though FL offered privacy advantages, it didn’t fully protect data, as sensitive information could still be figured out from the shared updates through attacks like membership inference. \cite{10.1145/3700838.3703679} This weakness showed the need to add more privacy-protecting methods to FL.

Federated Learning (FL)\cite{fed} was appreciated for the preservation of raw data on local devices, a significant leap towards more privacy \cite{mahlool2022comprehensivesurveyfederatedlearning}. Nevertheless, the observation that model updates were still indicative of sensitive information \cite{10.1145/3700838.3703679} demonstrated that FL itself was not adequate for robust privacy protection. This meant FL’s privacy benefits were limited and needed extra support. The takeaway is that FL should be seen as a tool that reduces privacy risks rather than one that fully ensures privacy, requiring combination with other privacy-protecting techniques for complete protection. FL’s decentralized approach naturally fit with data sovereignty principles and regulations like GDPR, which set strict rules on data transfers \cite{mahlool2022comprehensivesurveyfederatedlearning}. By keeping data local, FL avoided complications with cross-border data sharing. This made FL a key tool for AI in highly regulated fields. The larger point is that FL could be a practical framework for building compliant AI systems in dataspaces \cite{ds}, especially across different regions, as long as its privacy weaknesses are fixed.

\subsection{Differential Privacy (DP)}
Differential Privacy (DP) \cite{dps} was widely used in Federated Learning (FL) \cite{fed} to offer strong, math-based privacy protection by adding carefully controlled noise to data or model updates \cite{computers13110277}. This made it nearly impossible to tell if someone’s data was used in training \cite{AHMADZAI2023543}. DP could be applied during local training (protecting individual data points) or when combining updates from devices (protecting user data) in FL systems \cite{computers13110277}. However, standard DP methods used the same privacy level ($\epsilon$) for all clients, which often added too much noise and lowered model accuracy, especially when data differed greatly between clients (non-IID settings) \cite{AHMADZAI2023543}. This created a tough balance between protecting privacy and keeping the model useful \cite{mahlool2022comprehensivesurveyfederatedlearning}. Research showed that while DP improved privacy, it could unfairly affect smaller or underrepresented groups, raising fairness issues \cite{talaei2024adaptivedifferentialprivacyfederated}. Trying to make DP fairer often reduced accuracy, with bigger drops in non-IID settings \cite{10.1145/3700838.3703679}. Prioritizing more levels of fairness with Differential Privacy (DP) \cite{dps} often enlarges privacy threats, quantified by how much sensitive data might be approximated from attacks \cite{talaei2024adaptivedifferentialprivacyfederated}. As LDP was better at concealing specific client information, in general, it provided more protection of privacy than GDP \cite{9194237}.

A new approach to improve DP in Federated Learning (FL) \cite{fed} was suggested, using a Haar wavelet transformation and a new way to add noise \cite{computers13110277}. This method reduced the amount of noise needed by adding it to wavelet coefficients instead of directly to gradients and using per-level norm clipping \cite{s24227389}. This led to better model performance and faster training while keeping the same privacy protection \cite{computers13110277}. The research showed that Differential Privacy (DP) \cite{dps}, while offering strong privacy protection, created complex challenges with model accuracy and fairness \cite{talaei2024adaptivedifferentialprivacyfederated}. The finding that improving fairness could sometimes increase privacy risks \cite{Solove2025} pointed to a tricky problem involving multiple goals. This suggested that just using DP could unintentionally worsen issues like bias. The takeaway was that future DP studies should focus not only on balancing privacy and accuracy but also on addressing fairness and other ethical issues, possibly through flexible methods. A new DP approach using Haar wavelet transformation and a unique way of adding noise \cite{computers13110277} showed that advanced techniques could improve the balance between privacy and model performance. By working with transformed data and adjusting how noise is added, this method reduced noise while keeping the same privacy protection \cite{10.3389/fdata.2024.1497535}. This highlighted that a better understanding of data could lead to more effective privacy solutions. The larger point was that combining fields like cryptography, machine learning, and signal processing is essential for creating practical privacy-preserving AI systems.

\subsection{Trusted Execution Environments (TEEs)}
Trusted Execution Environments (TEEs) \cite{7345265} utilized hardware to enhance privacy by building secure, isolated environments for processing data \cite{Goretti_2024}. Data and code remained encrypted and inaccessible to the operating system or other software within a TEE, protecting them and keeping them secure \cite{10467913}. This allowed third parties or cloud services to process sensitive data without seeing the unencrypted information \cite{Goretti_2024}. Since TEEs could run regular smart contract programs without needing special languages or systems, they were more efficient and flexible than cryptographic methods like Secure Multi-Party Computation (SMC) \cite{smc} or Fully Homomorphic Encryption (FHE) \cite{Goretti_2024, 9734024}. TEEs enabled privacy regulations such as data being private by default, only the needed data being used, and restricting data use to their intended purposes, all without sharing data broadly but instead processing it securely \cite{9897507}. They also facilitated accountability with a process known as attestation, which showed code executed safely and complied with regulations \cite{ppc}. Potential uses include processing sensitive financial or medical data on blockchains, private AI agents, and dark pools \cite{Goretti_2024}. TEEs were not without limitations, though. If someone had direct access to the hardware, they might be open to sophisticated physical attacks, and no TEE was completely safe \cite{Goretti_2024}. Attackers could also track how data is accessed in the computer’s memory, so techniques like Oblivious RAM (ORAM) were needed to hide these patterns, though this slowed performance \cite{phe}. TEEs were not suitable for ensuring the integrity of blockchain systems but were suitable for preserving privacy \cite{Solove2025}. To counter these vulnerabilities, measures entailed preparing for potential security violations, prioritizing privacy rather than integrity, employing ORAM, and frequently updating security keys \cite{electronica}.

Trusted Execution Environments (TEEs) \cite{7345265} gave a big boost to performance by using hardware to create safe, private spaces for data processing \cite{Goretti_2024}. But they weren’t perfect, attackers could break in through physical attacks or by spotting patterns in how data was accessed in memory \cite{10568134}. This meant relying on TEEs required keeping the hardware safe and using additional software protections like Oblivious RAM (ORAM). Rather than fixing all security issues, TEEs shifted where the risks appeared. The main idea is that using TEEs for AI in dataspaces \cite{ds} called for a combination of defenses, such as pairing hardware security with encryption and cautious practices. TEEs worked faster than Fully Homomorphic Encryption (FHE) \cite{9734024} or Secure Multi-Party Computation (SMC) \cite{smc} for some tasks \cite{Goretti_2024}, so they helped close the gap where purely encryption-based methods slowed things down. Still, their weaknesses to physical attacks and the fact that ORAM could slow performance \cite{she} showed TEEs weren’t the answer for every privacy situation. Choosing the best privacy tool depended on the specific risks, how fast the system needed to be, and how much you could trust the hardware. In short, TEEs were great for cases needing good privacy and speed, as long as the hardware was kept secure or other steps reduced risks from sneaky attacks.

\subsection{Homomorphic Encryption (HE) and Secure Multi-Party Computation (SMC)}
Homomorphic Encryption (HE)\cite{homo} and Secure Multi-Party Computation (SMC)\cite{smc} were sophisticated cryptographic methods that supported computations on encrypted data or distributed inputs without compromising the underlying sensitive information. The approaches provided rigorous privacy assurances, especially applicable for extremely sensitive data in AI-driven dataspaces\cite{ds}.
Homomorphic Encryption (HE)\cite{homo} allowed calculations on encrypted data, with the decrypted result matching the operation on the original data \cite{10.1145/3715877}. This let cloud services or third parties process sensitive data, like analytics or AI tasks, without seeing the unencrypted information. Three categories of HE were identified: Partially Homomorphic Encryption (PHE)\cite{phe}, which was capable of performing a single calculation repeatedly; Fully Homomorphic Encryption (FHE)\cite{9734024}, which could execute any calculation, and Somewhat Homomorphic Encryption (SHE)\cite{she}, which could handle addition and multiplication for a finite number of times \cite{10.1145/3715877}. Although Fully Homomorphic Encryption (FHE)\cite{9734024} was the most powerful, it was too slow for large-scale business analytics and not practical for real-time uses \cite{Gamiz2025}. Homomorphic Encryption (HE) was particularly valuable for sensitive data in fields like healthcare or finance, where keeping data completely private was essential, meeting requirements of regulations like GDPR, HIPAA, and PCI DSS \cite{10.1145/3715877}.

Secure Multi-Party Computation (SMC) \cite{smc} let multiple groups collaborate on calculations without revealing their private data \cite{Gamiz2025}. SMC \cite{smc} used techniques like secret sharing or garbled circuits to make sure only the final result was shared, as long as most participants were trustworthy \cite{Gamiz2025}. It was perfect for situations where distrusting groups, like competing companies or healthcare providers, wanted to analyze data together without revealing their raw information \cite{10467913}. Compared to Fully Homomorphic Encryption (FHE) \cite{9734024}, SMC \cite{smc} was less demanding on computing power but needed a lot of communication between parties, which slowed things down in cloud settings, especially with more participants \cite{Gamiz2025}. There were important differences between Homomorphic Encryption (HE) \cite{homo} and SMC \cite{smc}. HE worked better when calculations were done in one place but struggled with complex data analytics tasks \cite{phe}, while SMC, on the other hand, was better for collaborative analytics across multiple groups but slowed down due to communication needs \cite{Gamiz2025}. For highly sensitive data, Homomorphic Encryption (HE) \cite{homo} provided stronger security, while Secure Multi-Party Computation (SMC) \cite{smc} was better suited for moderate-risk cases involving multiple parties \cite{Gamiz2025}. Research showed that both HE \cite{9734024} and SMC \cite{smc} managed privacy and fairness better than Differential Privacy (DP) \cite{dps}, with SMC \cite{smc} creating the smallest differences in outcomes across groups \cite{talaei2024adaptivedifferentialprivacyfederated}. However, both HE and SMC needed significant computing resources \cite{10.1145/3715877}. Using HE for encryption alongside SMC \cite{smc} for group calculations was considered a promising approach for handling large-scale, fast data analytics tasks \cite{Gamiz2025}.

Although Fully Homomorphic Encryption (FHE) \cite{9734024} and Secure Multi-Party Computation (SMC) \cite{smc} provide excellent privacy protection, they have significant drawbacks: FHE requires a lot of computing power, and SMC \cite{electronica} is slowed by the need for extensive communication between parties \cite{Gamiz2025}. This means that while these encryption methods work well in theory, they often face challenges in practical use, particularly for large AI projects in dataspaces \cite{ds}. The takeaway is that we need major improvements, like faster hardware (think GPU-powered FHE tools) or smarter ways to design these systems, to make them practical for more than just small, specialized uses \cite{Gamiz2025}. Comparing the two, HE and SMC \cite{smc} each have their own strengths and weaknesses, so they work better together than against each other \cite{Solove2025}. Homomorphic Encryption (HE) \cite{homo} is ideal to use on encrypted data in a single place, whereas Secure Multi-Party Computation (SMC) \cite{smc} is ideal when multiple parties work together without sharing data. It indicates the use of both together to overcome their respective shortcomings. The main concept is that the development of secure, privacy-aware AI systems for dataspaces \cite{ds} will most likely be based on strategically combining these approaches, leveraging each where it is best suited to optimize privacy and efficiency. The performance impacts and limitations of various privacy preserving techniques in dataspaces can be seen in Table \ref{tab:privacy_technologies}.

\begin{table*}[htbp]
\centering
\caption{Comparison of Privacy-Preserving Technologies in Dataspaces}
\label{tab:privacy_technologies}
\begin{tabular}{|p{2cm}|p{2cm}|p{2cm}|p{3cm}|p{2cm}|p{2.5cm}|}
\hline
\textbf{Technology} & \textbf{Approach} & \textbf{Privacy Guarantee} & \textbf{Performance Impact (Latency, Throughput, Computational Cost)} & \textbf{Scalability} & \textbf{Limitations} \\
\hline
Federated Learning (FL) & Decentralized model training; raw data remains local; only model updates shared \cite{mahlool2022comprehensivesurveyfederatedlearning} & Medium (mitigates raw data sharing, but vulnerable to inference from model updates) \cite{10.1145/3700838.3703679} & Low-Medium (reduces communication cost vs. centralized, but convergence can be slower) \cite{computers13110277} & High (scales to many clients/devices) \cite{mahlool2022comprehensivesurveyfederatedlearning} & Data leakage from model updates, non-IID data challenges, security across ecosystem \cite{10.1145/3700838.3703679} \\
\hline
Differential Privacy (DP) & Adds calibrated noise to data/updates; mathematically provable privacy \cite{computers13110277} & High (strong, provable privacy guarantees) \cite{AHMADZAI2023543} & Medium-High (accuracy degradation, especially with strong privacy or non-IID data; computational overhead for noise injection) \cite{AHMADZAI2023543} & High (can be applied to large datasets) & Utility-privacy trade-off, disproportionate impact on underrepresented groups (fairness concerns) \cite{talaei2024adaptivedifferentialprivacyfederated} \\
\hline
Trusted Execution Environments (TEEs) & Hardware-based isolated execution; code/data encrypted within enclave \cite{Goretti_2024} & High (confidentiality from OS/hypervisor) \cite{Goretti_2024} & Low-Medium (more efficient than FHE/SMC; overhead from ORAM if used) \cite{Goretti_2024} & Medium (limited by hardware availability/cost) & Vulnerable to physical attacks; side-channel attacks (memory access patterns); not for integrity \cite{Goretti_2024} \\
\hline
Homomorphic Encryption (HE) & Computations directly on encrypted data without decryption \cite{10.1145/3715877} & High (data remains encrypted end-to-end) \cite{10.1145/3715877} & High (FHE is computationally intensive, impractical for real-time BI; noise growth) \cite{Gamiz2025} & Low-Medium (challenges with large-scale BI analytics) \cite{Gamiz2025} & High computational cost, limited operational flexibility for complex queries, key management \cite{Gamiz2025} \\
\hline
Secure Multi-Party Computation (SMC) & Multiple parties jointly compute a function without revealing private inputs \cite{Gamiz2025} & High (no party learns anything beyond final output) \cite{Gamiz2025} & High (high communication overhead, especially with many parties; latency concerns) \cite{Gamiz2025} & Medium-High (more scalable for collaboration than HE, but communication limits) \cite{Gamiz2025} & High communication cost, vulnerable to collusion (depending on trust model), protocol complexity \cite{Gamiz2025} \\
\hline
\end{tabular}
\end{table*}

\section{Trade-off Strategies for Privacy and Performance}
Creating a balance between privacy and performance in AI-driven dataspaces \cite{ds} needed careful planning to handle the natural challenges. While Reinforcement Learning (RL) methods could adjust on the fly, this review focused on fixed and flexible strategies that didn’t rely on RL \cite{multi-big}.
\subsection{Static Approaches}
Static strategies for balancing privacy and performance used set privacy limits and fixed rules for compliance. In Differential Privacy (DP) \cite{dps}, traditional static methods applied the same privacy level ($\epsilon$) to all clients \cite{AHMADZAI2023543}. This was easy to set up and gave predictable privacy protection. However, it often added excessive noise, which reduced model accuracy, particularly when data differed significantly between clients (non-IID settings) \cite{mahlool2022comprehensivesurveyfederatedlearning}. The fixed noise level could either overprotect low-risk data, slowing performance, or underprotect high-risk data, creating privacy vulnerabilities \cite{talaei2024adaptivedifferentialprivacyfederated}. These static DP methods were simple to use and offered clear privacy benefits \cite{computers13110277}. However, because they lacked sufficient flexibility, their simple approach created issues in dataspaces with widely disparate data \cite{Solove2025}. Using the same privacy settings for everyone either put performance at risk by protecting data too much or put privacy at risk by protecting data too little. Static methods were a good place to start, but their shortcomings highlighted the need for more intelligent and adaptable strategies to better balance privacy and performance in practical AI projects.

\subsection{Adaptive (Non-RL) Strategies}
Adaptive non-Reinforcement Learning (non-RL) strategies aimed to customize privacy settings or pick suitable privacy tools depending on the context, without using reinforcement learning. Heuristic-based approaches used fixed rules or limits to adjust privacy settings. For example, in Differential Privacy\cite{dps} (DP), the amount of noise could be changed based on how sensitive or risky the data was. This gave more precise control than fixed privacy settings, aiming to keep data private without hurting performance too much. While the sources don’t directly call these methods “heuristic-based,” the idea of changing privacy levels based on things like “gradient sensitivity” \cite{AHMADZAI2023543} points to a rule-based approach to balance privacy and performance. Context-driven methods picked different privacy tools depending on factors like risk, performance needs, or the type of data. For instance, a system might switch between Differential Privacy\cite{dps} and Trusted Execution Environments (TEEs)\cite{7345265} based on a risk score, available computing power, or the data being handled. If a task needed high speed with less critical but sensitive data, a TEE might be used. For highly sensitive data spread across multiple places, combining Federated Learning\cite{fed} with Differential Privacy\cite{dps} could be better. Discussions about adjusting noise in DP \cite{AHMADZAI2023543} and making trade-offs based on context in Federated Learning systems \cite{talaei2024adaptivedifferentialprivacyfederated} support the idea of these flexible, situation-based strategies.

The drawbacks of fixed privacy settings \cite{AHMADZAI2023543} showed why adaptive strategies were so important. Ideas like "context-dependent trade-offs" \cite{talaei2024adaptivedifferentialprivacyfederated} and adjusting noise based on “gradient sensitivity” \cite{AHMADZAI2023543} made it clear that understanding the specific details of data, tasks, and the environment was key to getting the best balance between privacy and performance. This suggested that being more aware of the situation could lead to smarter adjustments without relying on Reinforcement Learning (RL). The takeaway was that future studies should work on building strong, real-time ways to sense the context and make decisions to guide privacy-preserving technologies (PETs) in fast-changing dataspaces. Although the focus was on non-Reinforcement Learning (non-RL) adaptive strategies, the sources frequently discussed adaptive Differential Privacy (DP)\cite{dps} using Reinforcement Learning (RL) to adjust privacy levels \cite{AHMADZAI2023543}. This highlighted a research gap or difficulty in identifying clear examples of non-RL adaptive methods beyond simple rule-based approaches. The complexity of fine-tuning privacy often led researchers to use RL. The bigger picture is that more research is needed into advanced non-RL adaptive methods, possibly borrowing ideas from control theory or decision-making frameworks, to offer simpler options for situations where RL might be too complicated or resource-heavy.

\subsection{Comparison with Dynamic RL-based Systems (Gap Identification)}
Reinforcement Learning (RL)-based systems could figure out the best balance between privacy and performance by constantly learning from their environment. They adjusted settings, like the privacy level ($\epsilon$) in Differential Privacy (DP)\cite{dps}, based on real-time information. \cite{AHMADZAI2023543} For example, one proposed system used an RL agent (based on PPO) in Federated Learning\cite{fed} to tweak privacy levels for each client, finding a sweet spot between keeping data private and maintaining model accuracy\cite{AHMADZAI2023543}. This method got results almost as good as non-private models, used less privacy budget, and trained faster than static DP approaches\cite{dps}. The big difference between non-RL adaptive strategies and RL-based systems comes down to how they learn and adapt. Non-Reinforcement Learning (non-RL) methods depend on static rules, plain instructions, or situational logic and are therefore simpler to implement and prove but less flexible or efficient. Reinforcement Learning (RL) systems instead learn the best methods through experimentation and error, tending to perform better but being more demanding in terms of computational resources, intricate training, and difficulty in explanation or guaranteeing reliability \cite{AHMADZAI2023543}. By leaving out RL from this review's scope, it identifies a lacuna in investigating novel non-RL techniques that might be as effective as RL without the overhead.

Reinforcement Learning (RL)-based adaptive strategies were great at dynamically tweaking settings to get the best balance between privacy and performance \cite{talaei2024adaptivedifferentialprivacyfederated}. But their "black-box" nature, common in complex AI systems like RL, made it hard to understand or prove how they managed the trade-off between privacy and performance. This suggested that while RL could adapt well, it might make things less clear. The takeaway for highly regulated dataspaces \cite{ds} was that RL’s flexibility had to be weighed against the need for systems that are easy to check, comply with rules, and explain clearly, which might make simpler, more transparent non-RL strategies a better choice when solid guarantees are critical. The survey’s choice to skip RL-based systems, even though sources praised their effectiveness \cite{10.3389/fdata.2024.1497535}, showed a clear focus on exploring other adaptive methods. This pointed to an awareness of RL’s challenges, like high computing needs, training data demands, and lack of clarity. RL’s limitations opened the door for new ideas in non-RL adaptive strategies. The bigger picture was a push for research into fresh approaches, like rule-based methods, optimization techniques, or control theory, that could draw inspiration from structured systems (like Fleming metadata) to achieve close-to-perfect dynamic balances without RL’s complexity, possibly using symbolic AI or formal methods to make things clearer and easier to verify. The comparison of Static and Adaptive Privacy-Performance Trade off strategies can be seen in Table \ref{tab:privacy_tradeoff_strategies}.

\begin{table*}[htbp]
\footnotesize 
\centering
\caption{Comparison of Static and Adaptive Privacy-Performance Trade-off Strategies}
\label{tab:privacy_tradeoff_strategies}
\begin{tabularx}{\textwidth}{|p{2.0cm}|X|X|X|X|X|} 
\toprule 
\textbf{Strategy Type} & \textbf{Characteristics} & \textbf{Advantages} & \textbf{Disadvantages} & \textbf{Applicability in Dataspaces} & \textbf{Distinction from RL-based Systems} \\
\midrule 
Static Privacy Budgets & Predefined, fixed privacy parameters (e.g., fixed $\epsilon$ in DP) applied uniformly across all data/clients. \cite{AHMADZAI2023543} & Simplicity in implementation and management; predictable privacy guarantees. & Suboptimal utility-privacy trade-off, especially in heterogeneous (non-IID) data environments; over-protection or under-protection for varying data sensitivities. \cite{AHMADZAI2023543} & Simple, low-variability dataspaces\cite{ds} where data characteristics are uniform and privacy requirements are constant. & Relies on fixed rules; no learning or dynamic adjustment based on real-time feedback. \\
\midrule 
Adaptive (Non-RL) Strategies & Dynamic adjustment of privacy parameters or selection of PETs based on predefined rules, heuristics, or explicit contextual factors (e.g., threshold-based noise tuning, context-driven selection). \cite{talaei2024adaptivedifferentialprivacyfederated} & More optimized trade-offs than static methods; can respond to varying data sensitivities or risk levels; potentially more interpretable than RL-based methods. & Requires careful design of heuristics/rules; may not achieve global optimality; can be complex to manage in highly dynamic environments if rules are extensive. & Dataspaces\cite{ds} with varying data sensitivities, dynamic workloads, or diverse user contexts where a balance between performance and privacy is critical, but RL complexity is undesirable. & Relies on pre-programmed logic or explicit contextual rules; does not learn optimal policies through trial-and-error interaction with the environment. \\
\bottomrule 
\end{tabularx}
\end{table*}

\section{Performance Impact Analysis and Benchmarking}
To figure out how well privacy-preserving and policy-aware AI systems work in dataspaces \cite{ds}, we need a solid way to measure their impact on performance. This means setting clear metrics and tackling the gaps in current benchmarking standards. \cite{multi-big}

\subsection{Key Performance Metrics}
The performance of AI systems in dataspaces\cite{ds}, especially when incorporating privacy-preserving mechanisms, was evaluated using the following key metrics:

\begin{itemize}
\item \textbf{Latency}: Indicates how long it takes a system to react to a request. It's crucial for AI tools that must function quickly and produce results right away.

\item \textbf{Throughput}: This indicates the speed at which a system can process information or tasks. It’s a way to see how much work the system can manage in a set amount of time.

\item \textbf{Cost Overhead}: This is about the extra computing power, communication, or energy needed when using privacy-protecting methods. For example, Fully Homomorphic Encryption (FHE) \cite{9734024} uses a lot of computing resources, while Secure Multi-Party Computation (SMC) \cite{smc} slows things down with heavy communication needs \cite{Gamiz2025}. Differential Privacy (DP) \cite{dps} also adds computing costs because of the noise it adds to data \cite{talaei2024adaptivedifferentialprivacyfederated}.

\item \textbf{Model Utility (e.g., Accuracy, F1-score)}: This checks how well an AI model does its job. Privacy methods often make models less accurate. For instance, adding more noise in Differential Privacy \cite{dps} usually lowers accuracy \cite{AHMADZAI2023543}. But newer methods, like utility-enhanced DP, have improved performance by cutting down on noise \cite{computers13110277}.

\item \textbf{Fairness}: Assesses the impartiality and fairness of an AI model's judgments for various groups, including those based on demographics. While Differential Privacy \cite{dps} promotes privacy, it can also have unfavorable effects on underrepresented or smaller groups, leading to concerns about fairness \cite{talaei2024adaptivedifferentialprivacyfederated}. On the other hand, Homomorphic Encryption (HE) and Secure Multi-Party Computation (SMC) \cite{smc} tend to balance privacy and fairness better, with SMC being especially good at keeping results fair across groups \cite{talaei2024adaptivedifferentialprivacyfederated}.

\item \textbf{Explainability/Interpretability}: This is about understanding how an AI model makes its decisions. It’s really important for checking if the system follows rules, especially for high-risk AI systems \cite{Enqvist2024}.

\item \textbf{Compliance Complexity}: This tracks how much effort and resources it takes to make sure the system follows relevant laws and policies.
\end{itemize}

In the past, folks thought AI performance was just about how accurate and fast it could be. But when you’re dealing with privacy-focused AI in dataspaces \cite{ds}, performance covers way more ground, like how quickly the system responds (latency), how much data it can process (throughput), the extra resources it eats up (cost overhead), and important stuff like fairness and explaining how the AI makes decisions. \cite{Gamiz2025} Including privacy shifts the focus from pursuing a single goal to managing multiple goals. This calls for a wider approach to evaluate AI systems, considering technical aspects, ethical issues, and legal requirements, beyond just accuracy. The research keeps showing that strong privacy tools, like Fully Homomorphic Encryption (FHE), Secure Multi-Party Computation (SMC), and Differential Privacy (DP) \cite{9734024, smc, dps}, come with big costs, think heavy computing demands, slower communication, or less accurate results. \cite{Gamiz2025} Privacy isn’t free; it takes a toll on performance. To make these tools work in real-world dataspaces \cite{ds}, you’ve got to weigh those costs carefully, which might mean making trade-offs or using special hardware to keep things practical. Cisco’s study also points out that getting these trade-offs right can be a big win for businesses. \cite{cisco2025}

\subsection{Current Benchmarking Landscape and Proposed Standards}
The current landscape of AI benchmarking was described as a "minefield," with a lack of standardization and concerns regarding what should be measured, according to what standards, and with what downstream effects.\cite{10.1145/3689598} While AI benchmarks played a central role in assessing model capabilities and risks, particularly in regulatory contexts like the EU AI Act \cite{Enqvist2024}, existing studies repeatedly showed weaknesses in benchmarks perceived as state-of-the-art.\cite{10.1145/3689598} There was a substantial need to scrutinize capability and safety-oriented AI benchmarks to the same extent as the AI models they were meant to evaluate, demanding transparency, fairness, and explainability from the benchmarks themselves.

Specifically for privacy-preserving AI, a significant lack of standardization existed for empirical privacy metrics for synthetic data generation\cite{electronica}. While various privacy metrics (PMs) had been proposed to assess privacy risk (e.g., Identical Match Share, Distance to Closest Record), there was no unified approach, and PMs often embodied the same assumptions as the mechanisms they assessed. This fragmented landscape made it challenging to compare different privacy-preserving AI systems objectively and to provide consistent guarantees across dataspaces\cite{ds}.

The research straight-up called evaluating AI a “minefield,” saying that the benchmarks we use need to be just as open, fair, and easy to understand as the AI systems themselves. \cite{10.1145/3689598} If people don’t trust the benchmarks, it messes up our ability to fairly judge how well AI systems work or follow rules. This means that when creating new standards for balancing privacy and performance in dataspaces \cite{ds}, we have to focus on making the benchmarking process itself reliable and easy to verify, not just the results. The EU AI Act’s push for benchmarks in high-risk AI systems \cite{Enqvist2024} shows that laws are demanding better, more consistent ways to test AI. This suggests that legal requirements are driving the need for stronger benchmarks. Looking forward, benchmarking won’t just come from academic interest or companies competing-it’ll be shaped by legal rules, which means tech experts, legal scholars, and policymakers need to team up to build and apply solid standards.

\section{AI Alignment Beyond Reinforcement Learning}
We need more than reinforcement learning to ensure that AI systems are morally well-behaved in data environments and aligned with human values. This involves the incorporation of easily interpretable logic, means for monitoring the behavior of the system, and simple methods for the enforcement of rules in AI processes.\cite{electronica}

\subsection{Symbolic AI for Compliance}
Symbolic AI, which uses clear reasoning and formats that people can easily read, showed great potential for improving AI compliance \cite{10.1145/3560819}. While traditional deep learning is excellent at spotting patterns, it often works like a hidden system that’s hard to understand. In contrast, symbolic AI gives logical results and can follow specific rules and expert knowledge \cite{RODGERS2023100925}. By combining symbolic AI with deep learning, known as neurosymbolic AI, we get a mixed approach that improves both clarity and prediction accuracy \cite{10.1145/3560819}. This blend uses deep learning to pick out important details and symbolic reasoning to make decisions that are easy to understand, providing clear explanations for AI choices \cite{barbara2023neurosymbolicaicompliancechecking}. For following regulations, neurosymbolic AI is essential because it helps auditors and regulators understand why AI makes certain decisions, especially in critical areas like financial risk assessment where clear reasons are needed \cite{PappilKothandapani2025}. It also reduces biases by setting fairness rules and aligning with ethical standards, supporting fair and responsible AI use \cite{SAURA2022101679}. By handling unclear data well and adapting better to new situations, neurosymbolic AI helps make accurate predictions and keeps up with changing regulations \cite{10.1145/3560819}. Deep learning models are often hard to understand, which conflicts with the need for clear explanations in high-risk AI, as required by regulations \cite{electronica, Enqvist2024}. Symbolic AI, with its clear and structured reasoning, provides the needed clarity \cite{10.1145/3560819}. The rise of neurosymbolic AI tackles this issue by blending the strengths of both approaches, offering easy-to-understand explanations while keeping strong prediction accuracy \cite{garcez2023neurosymbolic}. This shows that combining these AI methods is key to achieving both high performance and compliance that can be checked. More than just explaining decisions, the symbolic parts of neurosymbolic AI can actively enforce fairness rules and align AI results with set ethical standards \cite{feldman2025explainabilitycaseaivalidation}. This means symbolic logic can serve as a “guide” for AI behavior, reducing biases that might come from data. The larger point is that symbolic AI provides a strong way to build ethical considerations directly into the AI’s logic, moving from after-the-fact bias detection to proactive ethical alignment, which is essential for meeting regulatory requirements \cite{Solove2025}.

\subsection{Explainable AI (XAI) for Audit Trails}
Explainable AI (XAI) used methods like SHAP, LIME, and model-agnostic approaches to make AI systems easier to understand.\cite{electronica} The need for XAI was critical because unclear AI models often made biased or incorrect decisions, which reduced trust among users and stakeholders. Regulations, such as the EU AI Act, required AI systems to be clear and open to public review.\cite{Enqvist2024} A lack of openness was a key reason why people distrusted AI systems.\cite{Solove2025}
For tracking actions in data environments, XAI offered the clarity needed to see how AI systems handled sensitive data. This was essential for showing accountability and meeting data protection and ethical rules.\cite{9897507} Efforts were made to find the best ways to balance AI performance with clear explanations, building trust among developers, regulators, and users.\cite{electronica} In systems where data is spread out, like federated data ecosystems, making AI explainable for compliance was especially challenging.\cite{10.1145/3700838.3703679} Federated Learning\cite{fed}, which protects privacy, could still risk data leaks through model updates, so explaining these updates was vital to spotting potential weaknesses or biases.\cite{10.1145/3700838.3703679}
Although Explainable AI (XAI) was often seen as a tool to help people understand AI, its key role in meeting regulations and supporting audits was highlighted \cite{9897507}. The EU AI Act’s clear rules for explainability in high-risk systems \cite{Enqvist2024} showed that XAI was no longer just a research topic but a legal requirement for accountability. This meant XAI methods needed to improve to offer not only explanations for specific predictions but also broader insights into how models work and follow rules, suitable for official audits. In federated learning \cite{fed}, where data stays spread out to protect privacy, creating detailed audit trails and clear explanations was challenging because raw data was never in one place \cite{10.1145/3700838.3703679}. The risk of data leaks through model updates in federated learning \cite{talaei2024adaptivedifferentialprivacyfederated} made it even more important to explain these updates clearly. This showed that distributed learning systems made full explainability more complex. The larger point was that special XAI methods were needed for federated systems, focusing on explaining overall model behavior, the role of updates, and possible biases from scattered data, rather than just single predictions \cite{feldman2025explainabilitycaseaivalidation}.

\subsection{Policy Injection into AI Pipelines}
Rather than addressing compliance as an add-on, policy injection entailed incorporating legal and ethical regulations directly into various phases of AI development, ranging from data collection and preparation to training and deployment of the model. This meant building in requirements as firm rules or penalties to ensure AI systems naturally followed regulations, such as using only necessary data and sticking to specific purposes \cite{10.1145/3531532}. For example, before training, rules could be set to ensure only needed data was gathered and processed, with personal information properly hidden \cite{Solove2025}. During training, techniques could be used to enforce limits, like fairness rules or specific guidelines, by adding penalties to guide the AI toward compliant behavior \cite{RODGERS2023100925}. After deployment, ongoing monitoring and updating of AI policies were essential due to fast-changing laws \cite{10.1145/3531532}. This forward-thinking approach aimed to lower legal risks and encourage responsible AI use by making compliance a core part of the AI’s design and operation \cite{electronica}. Policy injection was not a one-time fix but a continuous effort across the entire AI process, from data collection to model use and monitoring \cite{SAURA2022101679}. This showed that inconsistent rule enforcement could create gaps in compliance at different stages \cite{PappilKothandapani2025}. To achieve consistent compliance, a single policy language or framework was needed to apply legal and ethical rules across all parts and stages of AI development, ensuring the system followed these rules throughout its life \cite{barbara2023neurosymbolicaicompliancechecking}. Legal frameworks like GDPR and the AI Act outlined broad principles and requirements \cite{Enqvist2024}. Policy injection provided a technical way to turn these general legal guidelines into specific, actionable rules within AI systems \cite{feldman2025explainabilitycaseaivalidation}. This showed that policy injection connected legal ideas to practical AI use. More broadly, effective AI governance relied on creating reliable tools and methods to convert legal text into policies that machines could follow, fitting smoothly into current AI development processes \cite{9897507}.

\section{Taxonomy and Comparative Analysis}
The main goal, degree of privacy protection, impact on performance, and complexity of compliance were the criteria used to create a clear and comprehensive system for classifying AI techniques that adhere to policies and protect privacy.\cite{10.1145/3531532} The purpose of this system was to provide a clear summary of the various options for AI-driven data environments as seen Table \ref{tab:taxonomy_ai_techniques_v2}.

\begin{table*}[htbp]
\centering
\caption{Taxonomy of Privacy-Preserving and Policy-Aware AI Techniques}
\label{tab:taxonomy_ai_techniques_v2}
\begin{tabular}{|p{3cm}|p{3.5cm}|p{2cm}|p{2.5cm}|p{3cm}|}
\hline
\textbf{Technique/Approach} & \textbf{Primary Focus} & \textbf{Privacy Level (Low/Medium/High)} & \textbf{Performance Degradation (Low/Medium/High)} & \textbf{Compliance Complexity} \\
\hline
Federated Learning (FL) & Data Decentralization, Collaborative Training & Medium (raw data stays local, but model updates can leak) \cite{10.1145/3700838.3703679} & Low-Medium (communication cost, convergence) \cite{computers13110277} & Medium (data quality, bias, threat vectors across ecosystem) \cite{10.1145/3700838.3703679} \\
\hline
Differential Privacy (DP) & Provable Privacy Guarantees & High (mathematically impossible to pinpoint individual data) \cite{AHMADZAI2023543} & Medium-High (accuracy degradation, especially non-IID) \cite{AHMADZAI2023543} & Medium (parameter tuning, fairness trade-offs) \cite{talaei2024adaptivedifferentialprivacyfederated} \\
\hline
Trusted Execution Environments (TEEs) & Hardware-based Isolation, Confidential Computing & High (code/data encrypted in enclave) \cite{Goretti_2024} & Low-Medium (faster than MPC/FHE, but ORAM overhead) \cite{Goretti_2024} & Medium (hardware limitations, physical attacks, trust assumptions) \cite{Goretti_2024} \\
\hline
Homomorphic Encryption (HE) & Computation on Encrypted Data & High (data remains encrypted end-to-end) \cite{10.1145/3715877} & High (FHE computationally intensive, impractical for real-time) \cite{Gamiz2025} & High (key management, scheme choice, noise growth) \cite{10.1145/3715877} \\
\hline
Secure Multi-Party Computation (SMC) & Joint Computation on Private Inputs & High (no party learns anything beyond output) \cite{Gamiz2025} & High (high communication overhead, scales with parties) \cite{Gamiz2025} & High (protocol complexity, trust assumptions, collusion vulnerability) \cite{Gamiz2025} \\
\hline
Constraint-based Optimization & Policy Enforcement during Training & Medium (depends on constraints, e.g., fairness, data use) \cite{RODGERS2023100925} & Low-Medium (optimization overhead, penalty adjustments) \cite{RODGERS2023100925} & Medium (formulating constraints, balancing objectives) \cite{RODGERS2023100925} \\
\hline
Rule-based Engines (for Governance) & Explicit Policy Enforcement, Audit Trails & Medium (depends on rules, data access) \cite{10.1145/3531532} & Low (for simple rules, but scalability for complex rules) \cite{PappilKothandapani2025} & Low-Medium (defining rules, managing rule sets) \cite{10.1145/3531532} \\
\hline
Symbolic AI / Neurosymbolic AI & Explainability, Logical Reasoning, Compliance & Medium (enforcing ethical/legal constraints) \cite{garcez2023neurosymbolic} & Medium (integration complexity, scalability for complex data) \cite{garcez2023neurosymbolic} & Low-Medium (human-readable explanations, auditability) \cite{garcez2023neurosymbolic} \\
\hline
Explainable AI (XAI) & Transparency, Interpretability & Low (indirect privacy benefit through transparency) \cite{electronica} & Low-Medium (computational cost of generating explanations) \cite{electronica} & Medium (regulatory demands, fostering trust) \cite{electronica} \\
\hline
Policy Injection & Embedding Legal Constraints & High (if enforced as hard constraints) \cite{10.1145/3531532} & Low-Medium (depends on implementation complexity) \cite{10.1145/3531532} & High (translating legal text to actionable policies) \cite{10.1145/3531532} \\
\hline
\end{tabular}
\end{table*}

\subsection{AI Explainability for Compliance in Federated Data Ecosystems}
Making AI explainable for compliance in federated data systems was a major challenge. Federated Learning (FL) \cite{fed} helped protect privacy by keeping raw data spread out, but it made creating clear audit trails and explanations for model decisions difficult \cite{10.1145/3700838.3703679}. Many AI models, especially deep learning ones, were already hard to understand \cite{electronica}. This problem grew worse in federated settings where data was distributed, and only model updates were shared \cite{10.1145/3700838.3703679}. The risk of data leaks through these model updates, even without accessing raw data, showed the need to explain the updates themselves to spot potential weaknesses or biases \cite{talaei2024adaptivedifferentialprivacyfederated}. Regulations like the EU AI Act clearly required explainability for high-risk AI systems \cite{Enqvist2024}. Connecting the decentralized setup of federated data systems with the centralized needs for compliance audits remained an unsolved issue. Developing explainable AI (XAI) methods that could clearly describe overall model behavior, individual contributions from clients, and possible biases in distributed data was essential for building trust and ensuring accountability in these complex systems \cite{10.1145/3700838.3703679, Enqvist2024}. The complexity of interpreting AI models, commonly referred to as a black box problem, made explainability difficult \cite{electronica}. Federated data systems introduced added complexity by maintaining data decentralized and only distributing model updates \cite{10.1145/3700838.3703679}. This demonstrated that distributed learning environments made it difficult to track model decisions back to individual data or contributions from clients, making the generation of detailed audit trails difficult \cite{feldman2025explainabilitycaseaivalidation}. It indicated that conventional explainable AI (XAI) approaches, designed for centralized models, did not work effectively in federated environments, necessitating novel methods capable of explaining decisions across distributed data sources and blended models \cite{9897507}. Although Federated Learning (FL) \cite{fed} was utilized to preserve privacy, the demand for explainability to satisfy compliance regulations could sometimes conflict with that privacy. For example, extremely precise descriptions of model updates could inadvertently spill information about what the underlying data is \cite{10.1145/3700838.3703679}. This referred to a problematic paradox: enhancing explainability for regulatory requirements in some instances might undermine privacy protections. The bigger picture was that there was a necessity for cautionary research into privacy-preserving explainability techniques for federated learning \cite{fed}, where explanations are designed so as to preserve privacy such that explaining decisions does not compromise the security of the data \cite{Solove2025}.

\subsection{Regulatory Gaps and Semantic Policy Enforcement}
Even with highly specific regulations such as the EU AI Act, difficulties remained in implementing policies in a practical manner across data environments enabled by AI. The primary challenge was translating into concise, machine-readable terms broad legal notions that could uniformly be applied across various technical systems and data processing flows\cite{10.1145/3531532}.
The international nature of regulatory environments was unbalanced, with varied data privacy and AI regulation approaches, causing ambiguity \cite{SAURA2022101679}. Although technology solutions like ODRL Data Spaces \cite{plazaortiz2025authenticationauthorizationdataspaces} and IDS RAM \cite{ids} aimed to standardize data sharing and compliance, challenges persisted due to differing global standards. The table directly compares different techniques in a clear side-by-side manner, so readers can easily see their key characteristics, advantages, and disadvantages \cite{10.1145/3531532}. The table enables researchers and practitioners to make properly informed decisions. By categorizing techniques according to performance impact and privacy level, the table shows the trade-offs addressed in the review. For instance, it indicates how high-privacy approaches like Homomorphic Encryption (HE) \cite{homo} and Secure Multi-party Computation (SMC) tend to bring notable performance losses \cite{10.1145/3531532}. The table also indicates where a method falters, such as HE's performance degradation, highlighting potential avenues of future improvement \cite{Gamiz2025}. Conversely, methods such as Federated Learning (FL) \cite{fed}, with their balanced trade-off of medium privacy and performance, shine through \cite{10.1145/3531532}. This juxtaposition forms a basis for the ensuing framework suggestion by establishing the major features of each method to build an integrated system \cite{Solove2025}.

\section{Research Challenges and Open Problems}
Even with notable progress, several key research challenges and unsolved issues remained in managing both privacy and performance in AI-driven data environments, especially when it came to aligning with policies and handling trade-off strategies.\cite{10.1145/3531532}

\subsection{Standardized Measurement of Privacy-Performance Trade-offs}
A major unresolved issue was the absence of standard methods and measures for evaluating the balance between privacy and performance in AI systems, particularly in data environments.\cite{electronica} Existing AI benchmarking practices were called a "minefield" due to the lack of agreed-upon standards for what to measure and how, raising ethical concerns and doubts about the reliability of current benchmarks.\cite{10.1145/3689598} For creating synthetic data that protects privacy, many privacy measures had been suggested, but without standardization, making fair comparisons challenging.\cite{electronica} This issue went beyond basic measures like accuracy and speed, involving the complex balance of privacy loss, model usefulness, computational demands, fairness, and explainability.\cite{Gamiz2025} Creating a clear, widely accepted set of measures to capture these varied trade-offs remained a significant challenge. Additionally, the political and performative aspects of benchmarks showed they were not neutral, requiring careful thought about the technical and ethical choices behind their design.\cite{10.1145/3689598}
The explicit description of AI benchmarking as a "minefield" and the lack of standardization for privacy metrics \cite{electronica} indicated a fundamental problem. Without consistent and trustworthy measurement, objective comparison of different solutions was impossible, hindering progress in the field and eroding trust among stakeholders. The implication was that establishing robust, transparent, and auditable benchmarking standards was not merely a technical task but a critical governance and trust-building endeavor for the entire AI dataspace ecosystem. The trade-off was not just between two metrics (privacy vs. performance) but involved multiple dimensions, including fairness and explainability.\cite{talaei2024adaptivedifferentialprivacyfederated} This complexity meant that "optimal" was context-dependent and involved ethical and societal considerations, not just technical ones. This indicated that the multi-faceted nature of the problem necessitated an interdisciplinary approach to defining and measuring trade-offs, involving computer scientists, ethicists, legal experts, and domain specialists. The broader implication was that future standardization efforts needed to incorporate multi-criteria decision-making frameworks and stakeholder engagement to define what constituted an acceptable balance in different application domains.

The clear labeling of AI benchmarking as a minefield \cite{10.1145/3689598} and the absence of standard privacy measures \cite{electronica} pointed to a core issue. Without reliable and consistent ways to measure performance, comparing different solutions fairly was impossible, slowing progress and reducing trust among users and stakeholders. This meant that defining clear, consistent, and auditable benchmarking standards was not just a technical exercise, but a key to building trust and good governance in AI data landscapes \cite{talaei2024adaptivedifferentialprivacyfederated}. The balance was not only between privacy and performance but also had to include aspects such as fairness and explainability \cite{feldman2025explainabilitycaseaivalidation}. This made it such that the best solution depended on the context and involved ethical and societal factors, not merely technical \cite{SAURA2022101679}. It highlighted the need for a multi-disciplinary process that brings computer scientists, ethicists, lawyers, and domain experts together in deciding and measuring these trade-offs. The big picture was that future standardization would have to involve templates for multi-factor considerations. Although technology solutions like ODRL Data Spaces \cite{plazaortiz2025authenticationauthorizationdataspaces} and IDS RAM \cite{ids} provided technical means to control data use, matching them with evolving legal documents, including new regulations in the AI Act, seemed complex \cite{plazaortiz2025authenticationauthorizationdataspaces}. The lack of one universal global AI regulation framework necessitated companies to implement flexible, compliant compliance measures \cite{zhao2025surveylargelanguagemodels}. Moreover, making AI systems inherently adhere to legal and ethical principles, as opposed to mere post-factum checks, demanded sophisticated policy integration methods and continuous monitoring \cite{10.1145/3531532}. The disparity between legible and understandable legal text and executable machine code made it difficult to enable fully automated and reliable policy execution in rapidly evolving data situations \cite{barbara2023neurosymbolicaicompliancechecking}. Legal frameworks like GDPR and the AI Act outlined what needed to be done, while technical tools like ODRL \cite{plazaortiz2025authenticationauthorizationdataspaces} and IDS RAM \cite{ids} explained how to do it, but turning complex legal ideas into clear, machine-readable rules remained a big challenge \cite{10.1145/3531532}. This showed that the difference between legal and technical language caused confusion in applying policies, risking compliance failures. It pointed to a strong need for formal methods, semantic web tools, and legal ontologies to connect these areas, allowing AI systems to automatically reason about and verify policy compliance. The fast-changing legal landscape around AI meant organizations had to keep updating their AI policies \cite{PappilKothandapani2025}. However, applying and updating policies across complex AI systems and scattered data environments was often slow and expensive. This indicated that laws could evolve quicker than organizations could adapt, leading to constant compliance challenges \cite{Solove2025}. The broader idea was the need for adaptable, "living" policy systems that could be swiftly updated and automatically implemented across AI systems in data environments, potentially using blockchain for secure policy updates or AI tools for automated policy interpretation and application \cite{9897507}.

\section{Future Directions}
The issues in balancing privacy and performance in AI-driven data environments called for several important future paths for research and development. These paths focused on creating a stronger, more compliant, and trustworthy AI ecosystem.\cite{multi-big}

\subsection{Conceptual Framework for Policy-Driven Alignment}
One crucial next step is to create a definitive model of how AI can be aligned with data environment policies. This model would incorporate legal and ethical principles into the AI system design, development, and deployment. It would offer an easy mechanism for translating general regulatory frameworks, such as those of GDPR and the AI Act, into machine-conformable rules via technical standards such as ODRL Data Spaces and IDS RAM \cite{plazaortiz2025authenticationauthorizationdataspaces}.

The framework would encompass:
\begin{itemize}
\item \textbf{Policy Specification Layer}: Employing formal languages, like ODRL extensions, to set clear rules, limits, and obligations for data usage and AI operations in data environments \cite{plazaortiz2025authenticationauthorizationdataspaces}.
\item \textbf{Policy Enforcement Layer}: Integrating tools such as constraint-based optimization and rule-based systems into AI workflows to maintain compliance in real time during training, decision-making, and data sharing \cite{10.1145/3531532}.
\item \textbf{Trust and Verification Layer}: Incorporating features for proof, auditability, and explainability (using XAI and Symbolic AI) to offer clear evidence of compliance and transparency in AI decision-making \cite{9897507}.
\item \textbf{Adaptive Governance Layer}: Creating systems for continuous monitoring, risk assessment, and flexible policy updates based on changing regulations and contexts, possibly using non-reinforcement learning approaches \cite{10.1145/3531532}.
\item \textbf{Interoperability Layer}: Ensuring smooth interaction and data sharing across different data environments and cloud systems while adhering to standard protocols \cite{dam}.
\end{itemize}

The review underscored that aligning AI systems is more than technical, it's directly related to legal and ethical concerns.\cite{10.1145/3531532} A clear framework was necessary to demonstrate how these fields are interconnected. By consolidating these various fields into a single approach, the framework could enable responsible AI development. This implies that the next step in research should be to develop models and ontologies that extensively explain the relationships between legal norms, moral values, and technological devices, enabling AI systems to reason on compliance and ethical conduct automatically.

\subsection{Automated Compliance Validation and AI-Driven Explainability}
Improvements in automated compliance validation were vital for growing policy-aware AI in data environments. This required building tools and methods to automatically confirm that AI systems follow specific policies and regulations, minimizing manual work and errors \cite{PappilKothandapani2025}.

\begin{itemize}
\item \textbf{Compliance Workflows Powered by AI}: Leverage AI, machine learning, and automation to simplify regulatory processes, detect risks, and ensure compliance using auto-rule checks, intelligent alerts, and automated audit reports.
\item \textbf{Policy Enforcement through Formal Verification}: Applying mathematical techniques to verify that AI systems adhere to their defined policies, particularly in high-risk environments. This would render AI systems more reliable and dependable in critical domains.
\item \textbf{AI-Driven Explainability}: Creating advanced explainable AI (XAI) techniques to produce clear, human-readable explanations for complex AI decisions, particularly in distributed and federated systems \cite{electronica}. This includes using neurosymbolic AI to provide clear reasoning and transparent decision-making processes \cite{10.1145/3560819}. The aim is to go beyond just offering explanations by automating the audit trail, allowing AI systems to report their compliance status and decision reasons in a format that can be easily checked.
\end{itemize}

The challenge of managing and monitoring AI compliance was becoming more complex. However, AI itself provided ways to automate these tasks, such as using AI-powered compliance workflows and AI-driven explainability \cite{PappilKothandapani2025, electronica}. This showed that AI could help solve the governance issues it caused, making compliance more efficient and accurate. This points to the need for "meta-AI" research, where AI systems are built to oversee, explain, and ensure the compliance of other AI systems, creating a self-monitoring and self-checking system.

\subsection{Standardization for Privacy-Performance KPIs}
Creating standard Key Performance Indicators (KPIs) for balancing privacy and performance was crucial for fairly evaluating and comparing AI systems in data environments. This would tackle the existing problem of inconsistent privacy metrics and benchmarking \cite{electronica}.

\begin{itemize}
\item \textbf{Multi-dimensional KPIs}: Creating KPIs that measure not just standard performance factors like speed and efficiency but also clear measures of privacy loss (such as understandable $\epsilon$ values and attack success rates), fairness, and explainability scores.
\item \textbf{Industry-wide Adoption}: Encouraging research, industry, and regulatory groups to use these standardized KPIs to ensure consistent evaluations and support the development of reliable AI solutions.
\item \textbf{Benchmarking Frameworks}: Developing open-source tools and datasets for consistent and repeatable testing of AI systems under various privacy-protecting configurations and policy rules. This would enable organizations to see the practical impacts of different trade-off approaches.
\end{itemize}

The lack of standard KPIs for privacy-performance trade-offs made objective comparison and incentivizing the creation of genuinely balanced solutions challenging. This meant that understandable, standardized KPIs would be a common language to use when assessing and optimizing AI systems, hence pushing innovation towards more compliant and responsible designs. The implication was that the creation of these KPIs was not merely a technical process but a strategic necessity for defining the future direction of AI research and deployment in dataspaces\cite{ds}.

\subsection{Integration with European Initiatives (GAIA-X, IDS, Eclipse EDC)}
Grounding the survey in practice involved close integration with ongoing European initiatives that aimed to build foundational data ecosystems.

\begin{itemize}
\item \textbf{GAIA-X}: This project aimed to create a blueprint for a federated data system based on such values as data sovereignty, privacy, confidentiality, and interoperability \cite{dam}. Future research must investigate how AI systems with privacy-protecing mechanisms and policy-congruent designs can be deployed and tested within the GAIA-X ecosystem\cite{gaia}, taking advantage of its compliance services and federated architecture \cite{dam}.

\item \textbf{International Data Spaces (IDS)}: IDS\cite{ids} aimed to foster trust, ensure security and data sovereignty, and support standardized interoperability in data ecosystems \cite{dam}. The IDS\cite{ids} Reference Architecture Model (RAM) and its usage control policies offered an essential technical base \cite{plazaortiz2025authenticationauthorizationdataspaces}. Future research should examine how AI models can directly adopt and apply IDS\cite{ids} usage control policies to uphold data sovereignty throughout the AI lifecycle.

\item \textbf{Eclipse Dataspace Components (EDC)}: Built on Gaia-X and IDSA standards, EDC\cite{arnold2025servicearchitecturedataspaces} provided key components (Connectors, Federated Catalog, Identity Hub) for secure and controlled data sharing in dataspaces \cite{dam}. Future work should concentrate on embedding privacy-preserving AI models and policy enforcement tools into EDC-based dataspace systems\cite{arnold2025servicearchitecturedataspaces}, using them as a testing ground for real-world deployment and validation of proposed solutions. Projects like Eona-X and Catena-X, which already utilize EDC\cite{arnold2025servicearchitecturedataspaces}, offer practical settings for such integration \cite{dam}.
\end{itemize}

Initiatives such as GAIA-X\cite{gaia}, International Data Spaces (IDS)\cite{ids}, and Eclipse Dataspace Connector (EDC)\cite{arnold2025servicearchitecturedataspaces} were not concepts but actual implementations developing data infrastructure \cite{dam}. These initiatives provided tangible environments to develop, test, and verify AI solutions that ensure privacy and abide by established rules. This implies that educational research needs to collaborate closely with such projects, utilizing them as actual test cases whereby theory can be applied to practical applications to facilitate the acceleration of the development of reliable AI in data-sharing contexts.

\section{Conclusion}
In the field of AI adoption in Dataspaces, this study shows the challenges of balancing between privacy and performance. Advancements in policy-aware AI alignment have been demonstrated by technical policy languages like ODRL Data Spaces\cite{plazaortiz2025authenticationauthorizationdataspaces} and IDS RAM, as well as regulatory frameworks like the EU AI Act and GDPR. By using strategies like constraint-based optimisation and rule-based systems, these frameworks enable proactive policy enforcement. Privacy-preserving techniques like Federated Learning, Differential Privacy, Homomorphic Encryption, Trusted Execution Environments, and Secure Multi-Party Computation provide a range of data confidentiality solutions. Custom hybrid architectures are required because each of these strategies has distinct trade-offs in terms of performance, scalability, and privacy guarantees. The lack of standardised metrics for assessing privacy-performance trade-offs, the difficulty of achieving explainability in federated data ecosystems, and the challenge of converting legal principles into machine-executable policies are enduring obstacles, though. To bridge theoretical advancements with real-world deployment, future research should focus on creating a unified framework for policy-driven AI alignment, advancing automated compliance validation, establishing standardised KPIs, and integrating with European initiatives such as GAIA-X, IDS, and Eclipse EDC. These efforts will ensure dataspaces respect fundamental rights and foster public trust by fostering transparent, auditable, and ethically aligned AI systems through interdisciplinary collaboration.

\end{document}